# Erwin Schrödinger: His Poetry and Fragment of a Galilean Dialogue

Helge Kragh*

**Abstract**. While Erwin Schrödinger's contributions to quantum mechanics and other areas of theoretical physics are very well known, his works as a poet and literary writer are usually ignored. In this paper I reconsider his poetry and refer to a few poems written about him. A short fragment of a Galileo-inspired dialogue that Schrödinger published in 1955 is as interesting as it is puzzling and little known. The dialogue is essentially about atomic theory as known much earlier, namely in the early 1910s. Indeed, there is the slight possibility that Schrödinger may originally have conceived it at that time and even composed a version of it. Apart from considering the origin and reception of the manuscript, I analyze its content from the point of view of the history of physics.

## Introduction

Erwin Schrödinger is principally celebrated for his fundamental contributions to quantum physics as epitomized by the eponymous wave equation that first appeared on p. 361 in volume 79 of the *Annalen der Physik*. The article was dated January 27, 1926 and published one and a half month later. To philosophers and the broader public he is associated with the famous cat-in-the-box thought experiment which he introduced in *Die Naturwissenschaften* in 1935. Incidentally, this article was largely ignored by contemporary physicists and only became widely cited many years after Schrödinger's death (see the figure).

Apart from being a brilliant physicist, Schrödinger was also a humanist scholar, a polymath who engaged in cultural studies relating to music, theatre, philosophy, history, philology, and literature [Gronau and Gronau 2021]. *Science and Humanism* and *Nature and the Greeks*, both books written while he stayed in Dublin, bear witness to his broad cultural interests. In this note, I briefly comment on Schrödinger's poetic contributions and then, in greater detail, on an interesting but little known dialogical fragment in which he discussed some aspects of atomic theory well before the quantum revolution.

* Niels Bohr Institute, University of Copenhagen, 2200 Copenhagen, Denmark. E-mail: helge.kragh@nbi.ku.dk.

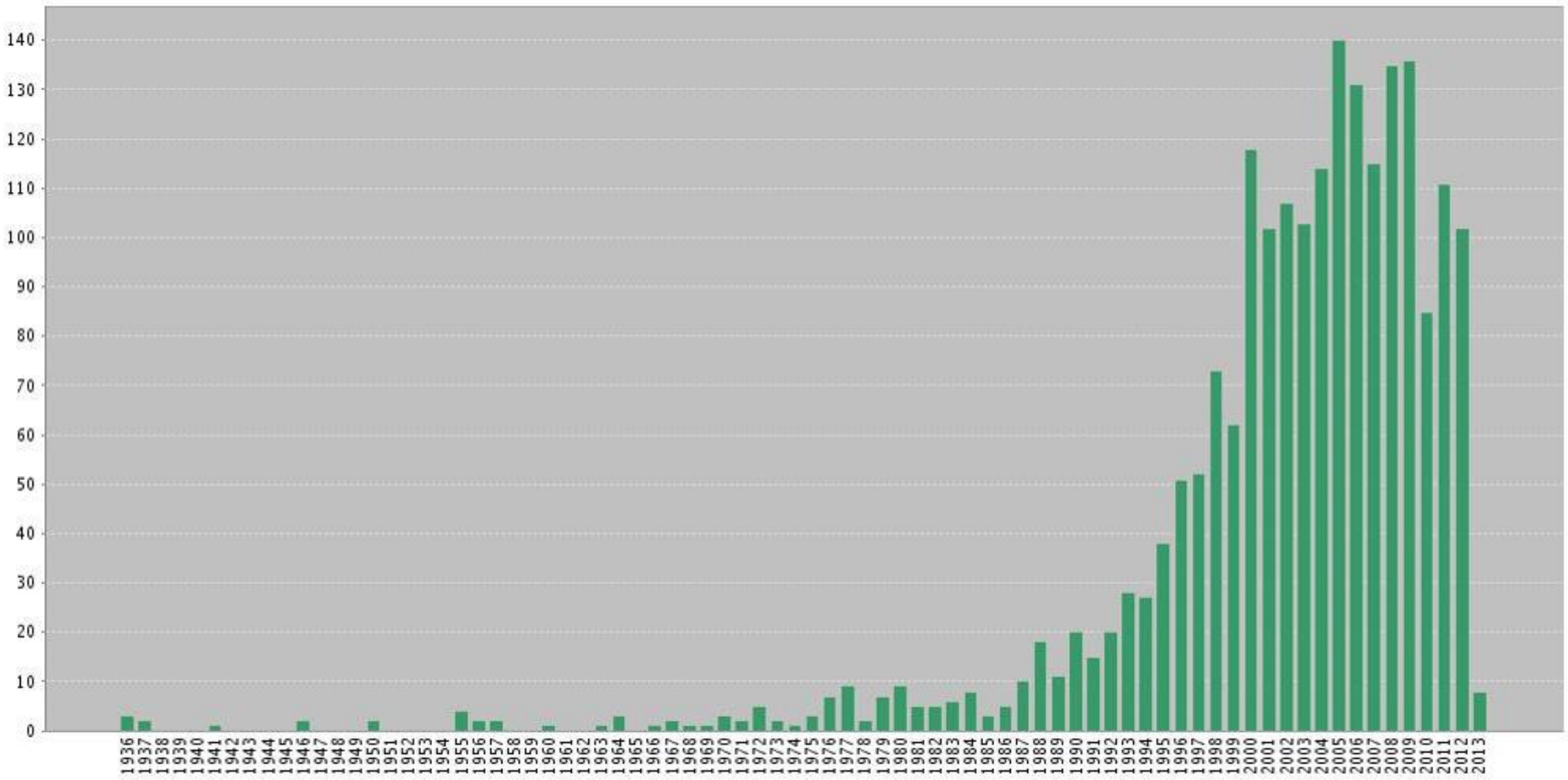

Number of citations to Schrödinger's cat-in-the box article from 1935 based on data from Web of Science.

## Schrödinger's poetry

Felix Bloch, a winner of the 1952 physics Nobel Prize, once recalled a conference held in Zurich in the summer of 1926, just a few months after Schrödinger had introduced the ψ function and the equation governing it. The young physicists attending the conference were incited to compose verses on their professors, which resulted in this one on Schrödinger [Bloch 1976, his translation from German]:

> Erwin with his psi can do
> Calculations quite a few.
> But one thing has not been seen:
> Just what does psi really mean?

Since then, numerous poems have been written on the father of wave mechanics, many of them dealing with the legendary half-dead cat. Schrödinger was himself a prolific writer of poetry, an art which filled much in his life. Thus, in a 1931 interview by the English journalist and science writer John W. N. Sullivan he said: "I took up mathematics and physics quite early … But I must not give the impression that science alone interested me. As a matter of fact, my early desire was to be a poet. But I speedily realised that poetry was not a paying business. Science, on the other hand, offered me a career" [Sullivan 1931; Sullivan 1934, p. 154]. The interview was reproduced in *Contemporary Mind*, a 1934 collection of essays based on Sullivan's conversations with famous scientists including (apart from Schrödinger) Arthur

Eddington, James Jeans and Max Planck, as well as two prominent British men of literature, namely Aldous Huxley and H. G. Wells.

Throughout his life Schrödinger wrote numerous poems, most of them occasional and for private consumption only [Sofronieva 2014]. A large part of them were poems related to his many love affairs. As an example, after a swim in a lake near Zurich, he wrote a sonnet. In the translation of Walter Moore [1989, p. 151], two of the stanzas read as follows:

> The brooding sun rests softly on the lake
> And barely marks its shallow breathing
> As gentle waves rock up and down
> In heavy glare of midday heat.
>
> Your eyes then close in peaceful satisfaction
> What comfort now to feel the loving water
> Embrace caressingly your waist and limbs
> And sweetly cool the heated blood.

In 1949, Schrödinger published a small collection of his poems, called *Gedichte*, most of which were in German but some in English [Schrödinger 1949; Janés 2022, pp. 177-270]. In one of the poems, he expresses how his work as a physicist is characterized by a religious attitude to nature. In the translation of his former student Bruno Bertotti [1985] it reads:

> Holy one, before you I kneel
> through you I draw
> the breath of the world.
> I am yours.
> When it pleases you
> I cease to be.

Schrödinger's poems were about love, life, nature, emotions, and his inner feelings, whereas none of them referred to physics or other sciences, at least not in any direct and recognizably sense. While some scientists writing poetry have referred to scientific themes in their verses, others have kept science apart. The noted French astrophysicist Jean-Pierre Luminet belongs to the latter category. "My poetic works had nothing to do with astronomy," he states. "For me, poetry had to express feelings, emotions, and subjects that cannot be reached by rational investigation, such as love, death, beauty, loneliness, despair, and so on" [Luminet 2015]. It seems that Schrödinger shared Luminet's attitude to poetry.

Terms such as 'physics,' 'atom,' 'electron' and 'quantum' are conspicuously absent from Schrödinger's *Gedichte* and other of his poems. As far as I can see, there is no clear connection between the poems and Schrödinger's ideas about physics and science generally. Nonetheless, according to Tzveta Sofronieva [2014], Schrödinger's work as a poet is "closely related to his ideas in science." Without providing concrete examples, she maintains that the poems illustrate "the interplay between literary imagination and scientific investigation." Schrödinger's late debut as a poet was not well received in literary circles and soon forgotten. The famous Austrian writer Stefan Zweig, the author of *The Royal Game* (German: *Schachnovelle*), is to have told him, "I hope your physics is better than your poetry." Zweig's comment is reproduced in some scholarly works but never provided with a source [e.g. Moore 1989, p. 225 and Middleton 2015, p. 66]. I suspect that the story is anecdotal.

## Origin and reception of Schrödinger's *Fragment*

American-born Dick Ahlstrom was a respected science editor of *The Irish Times* and since 2010 a member of the Royal Dublin Society. On April 18, 2012, he reported in the newspaper that one of Schrödinger's manuscripts had now resurfaced after a period of nearly sixty years [Ahlstrom 2012]. The five-page manuscript in question was titled "Fragment from an Unpublished Dialogue of Galileo" with Schrödinger adding after the title "Time: Between about 1910 and 1920."

Ahlstrom said about the manuscript that although it was "about the start of quantum mechanics in the early 1900s," it was placed as a dialogue with speakers "from the 16th century [*sic*] and a time when none of the ideas would have made any sense." Without referring to Galileo, Ahlstrom stated: "The character Salviati attempts to explain the oddities associated with quantum mechanics, while Simplicio refuses to believe that such things could possibly be true." According to the journalist, Schrödinger wrote the fragmentary manuscript in 1955, shortly before returning from Dublin to Vienna. He offered the typewritten copy to his friend Ronald Anderson, who taught English at the King's Hospital School and served as editor of the school's *Blue Coat* magazine [Moore 1989, p. 371]. Anderson then arranged that it was published in an issue of the informal magazine from December 1955 [Schrödinger 1955]. The manuscript was first owned by Anderson and later by other persons until it was handed over to the King's Hospital School.

Apparently unknown to Ahlstrom in 2012, the *Fragment* (as I shall call the manuscript) was republished twenty years later in the peer-reviewed academic journal *Hermathena* devoted to classical studies and published by Trinity College

Dublin [Schrödinger 1975]. The journal acknowledged Anderson for handing over the manuscript and also "Frau von Braunizer, daughter of Dr Schrödinger" for her permission to publish it. The reference is to Ruth March-Braunizer, the daughter of Schrödinger and his mistress Hildegunde March, who was the wife of Schrödinger's close friend, the Austrian theoretical physicist Arthur March [Moore 1989]. (After March's death in 1957, Ruth married Arnulf Braunizer). At the time of Ahlstrom's note in *Irish Times*, Schrödinger's *Fragment* had thus been published twice and was readily available to interested scholars.

The two versions of 1955 and 1975 are identical. A third unpublished version, unfortunately undated, is available from the University of Vienna archives, where it appears together with a letter from 1951 [Vienna]. Apart from a few misspellings and some stylistic differences, it is identical to the published 1955 version of which it seems to be a slightly earlier draft. On this occasion Schrödinger also added a German translation of the first half of his draft.

Despite being published, the *Fragment* piece remained curiously invisible for a long time and still today it has attracted very little scholarly attention. Thus, it is not included or even mentioned in Schrödinger's four-volume *Gesammelte Abhandlungen* [Schrödinger 1984] and nor does it appear in Walter Moore's comprehensive biography. There is also no trace of the *Fragment* in Schrödinger's published correspondence [Meyenn 2011] or in other sources that I know of. One might believe that scholars of the histories of physics and literature had paid attention to the printed manuscript, but with one notable exception this seems not to have been the case. Schrödinger passed away in early 1961 without ever referring to his *Fragment*.

The exception is a little known paper in Spanish from 2017 in which the author, the noted writer and poet Clara Janés, compared Schrödinger's *Fragment* and his general philosophy of science with Galileo's famous *Dialogo* (*Dialogue Concerning the Two Chief World Systems*) from 1632 [Janés 2017]. She reprinted her paper in a book of 2022 which also includes a Spanish translation of the *Fragment* and translations of the poems in Schrödinger's *Gedichte* [Janés 2022]. Unfortunately, neither Janés' paper nor book have not been translated into English or German, and probably for this reason her paper remains undeservedly unknown to historians of physics.

Schrödinger explicitly framed his fragment as a pastiche of Galileo's celebrated work, which he evidently admired and knew by heart. The same was the case with his contemporary Albert Einstein, who in 1918 adopted a somewhat similar framework in a playful Galilean dialogue on the theory of relativity published in the widely read journal *Die Naturwissenschaften* [Einstein 1918; Rowe and Schulmann 2007, pp. 96-103]. But whereas Schrödinger kept to Galileo's three characters (Sagredo, Salviati, Simplicio), Einstein chose the names 'Relativist' and 'Kritikus' for his two discussants. Many years later, Einstein wrote an insightful essay on Galileo as a foreword to Stillman Drake's English translation of *Dialogo* [Einstein 1953].

When did Schrödinger write his *Fragment*? Since he delivered the manuscript to Anderson in the autumn of 1955 it has been taken for granted that this was also the year in which he composed it. By that time Schrödinger had for long been occupied with conceptual and other problems of quantum physics, which makes it tempting to read elements of quantum mechanics into the *Fragment*, such as Ahlstrom did in 2012. Sofronieva [2014] likewise claims that the manuscript echoes the author's "deep involvement … with the discussions on the interpretation of quantum mechanics." And Janés [2017] finds "quantum uncertainty" hidden in Schrödinger's fragment. However, as reproduced in *Hermathena* the fragment refers exclusively to terms and concepts of classical physics. There is not a single allusion to quantum theory (and also none to the theory of relativity). This may present a puzzle, for what reason could Schrödinger possibly have had in 1955 to write a Galilean pastiche referring *only* to physics as known at about 1914?

Just a few years before 1955, Schrödinger had launched a broad attack on the Bohr-Heisenberg notion of discrete quantum jumps in the pages of the *British Journal for the Philosophy of Science*, introducing the lengthy article with a quotation (in Italian!) from Galileo's *Dialogo* [Schrödinger 1952]. When asked by Anderson to write a piece for the *Blue Coat* magazine, it would have been natural for him to turn his disagreement with Bohr, and generally with the Copenhagen interpretation of quantum mechanics, into a Galilean dialogue. This would presumably have been appreciated by readers of the magazine. Or he could have chosen some other topic which interested him at the time, say cosmology or the physical basis of life. But Schrödinger instead presented a fragmentary dialogue totally unrelated to the philosophical questions of the quantum world, a story which must have seemed strangely irrelevant to readers of the magazine.

The mentioned puzzle disappears if we assume, somewhat arbitrarily, that Schrödinger did not originally write the fragment in 1955 but at some time between 1911 and 1915, the period to which it refers. If so, the manuscript can have nothing to do with the theory of quantum mechanics developed more than a decade later. Schrödinger only took up quantum and relativity theories at about 1920. *Fragment* is about something else and less glamorous, reflecting physics as known by Schrödinger at about the outbreak of World War I. One could imagine that he wrote it down at the time, presumably just as an innocent pastime and with no intention of publishing it. If this is what he did, the original piece would have been in German.

To continue the speculation, when Anderson asked him for a contribution to *Blue Coat* some forty years later, Schrödinger may have decided to dust off his old fragmentary manuscript and rewrite it into an English typewritten version. He may have added the subtitle ("Time: Between about 1910 and 1920") on that occasion.

Admittedly, there is no evidence, archival or otherwise, that the manuscript dates from about 1914, which weakens the hypothesis very considerably. On the other hand, there is also no incontrovertible evidence that it *originally* dates from 1955. The first option is speculative and perhaps not very plausible, but at least it offers an explanation of the content of the Fragment, which the 1955 assumption does not. It is possible that an examination of the large amount of material in the Schrödinger archive, including notebooks and letters, can cast some light over the question.

## Physics and atomic theory in the early 1910s

Schrödinger, who received his doctorate from the University of Vienna in 1910, started his career with works on electricity, magnetism, kinetic theory, X-rays, and atmospheric radioactivity [Moore 1989, pp. 49-78; Mehra and Rechenberg 1987, pp. 64-228]. His doctoral thesis was a detailed examination of the conductivity of insulators. One of the questions that occupied him was the nature of electricity in terms of discrete quanta (electrons) and the relation of electricity to the structure of matter. While Galileo's *Dialogo* was about models of the universe, the Copernican world picture versus the traditional one of Aristotle and Ptolemy, Schrödinger's pastiche was concerned with the microscopic structure of matter.

The brief *Fragment* manuscript starts with the knowledgeable layman Sagredo introducing the topic of discussion, namely the claim that "All the ultimate building stones composing matter are … electrically charged." This, he says, follows from "the theory of our academician at Leiden and from the various experiments devised to prove it." As pointed out in Janés [2017], there is little doubt that Schrödinger is here referring to Hendrik A. Lorentz, the eminent theoretical physicist at Leiden University who won a Nobel Prize in 1902 and whose authoritative book *The Theory of Electrons* was published seven years later. The progressive Salviati (who in Galileo's book defends the Copernican system) agrees with Sagredo's claim. He proceeds to describe what is in effect Robert Millikan's famous experiments with charged oil drops conducted in the years 1910-1911:

> [When] a minute oildrop, just detectable in the microscope … catches one single elementary particle and its charge, you will see that the droplet at once starts wandering, provided that you have procured to establish an electric field in the chamber in which the droplet is suspended; and if subsequently it catches two or three of these charges, it wanders twice or thrice.

Now the conservative sceptic Simplico intervenes, which results in a brief methodological discussion which in spirit is close to some of the discussions appearing in Galileo's *Dialogo*:

> SIMPLICIO: Whence is it known *when* it catches a charge?
> SALVIATI: Well, just from the fact that it starts wandering or changes its velocity.
> SIMPLICIO: Surely you are not serious in this, Sr Salviati. Is not this the kind of inference which Aristotle has branded as a vicious circle?
> SALVIATI: Allow me to interpose a question: when you left us yesterday night and went home, did it rain?
> SIMPLICIO: Indeed it did, and you know it. You got wet on the short way to the garden gate, whither you accompanied us.
> SALVIATI: And whence knew you that it was raining?
> SIMPLICIO: Pardon me, but I am amazed at these naive questions of yours: of course from the fact that we got drenched.
> SALVIATI: So you would say: if it rains, one gets wet?
> SIMPLICIO (angrily): Indeed, just so.
> SALVIATI: Would not Aristotle have termed this a vicious circle, since we have just stated: the fact that it rains is noticed from the fact that one gets wet?

To Sagredo's question concerning the electric tension on a charged elementary particle, Salviati replies that it is "something like 2,000 electro-static units or, in round terms, half a million volts," a surprisingly large value. The figure corresponds to the rest mass of an electron as given by its energy, $m = 0.511\ \mathrm{MeV}/c^2$. SAGREDO: "Are we to understand that all and every matter consists of such highly charged particles? Even the water in the river, even the rain drops?" Salviati confirms, adding that this is also the case with the particles making up our own bodies. "Why should they make an exception? They too are, if you please, charged to about half a million volts."

The conversation soon turns to atomic theory and specifically to Salviati's surprising claim that matter as composed of atoms and molecules is almost completely empty. SAGREDO: "This, you said, has been proved by most ingenious experiments of that great natural philosopher at Cantabriga [Cambridge]." The name of the great natural philosopher is not given in the text, but it is obviously Ernest Rutherford:

> Salviati: Quite so. He shot swift corpuscles through the atoms and found that the body of the atom is mainly 'air' (you know what I mean!). The flying corpuscle is deviated only on the rare occasions when it passes very near to an atomic nucleus. And this nucleus is a 100,000 times smaller than the atom! However, the interstices are not really quite empty. They are the playground of electromagnetic fields, much stronger than any we are able to produce in the laboratory.

Salviati (Schrödinger, of course) refers to the ground-breaking scattering experiments with alpha particles made from 1906 to 1913 in the Cavendish Laboratory by Hans Geiger and Ernest Marsden. (The latter is misspelled 'Mardson' and 'Mardsen' in Janés [2017]). It was from these experiments that Rutherford deduced the existence of a tiny atomic nucleus, although he did not use that name for what he called the "central charge." In his 1911 landmark paper in *Philosophical Magazine*, he estimated the radius of a heavy nucleus to be approximately $10^{-14}$ m. The nuclear model aroused at first little attention and only became generally known after about 1913, when Bohr extended it to an atomic model based on quantum theory. There is no hint in Schrödinger's manuscript of either the Bohr model or of quantum theory.

The dialogue fragment ends with a discussion of light rays as electromagnetic fields and their relation to the fields within the atom which according to Salviati are "much stronger than any we can produce in the laboratory." Simplicio now argues that because light consists of the much weaker electromagnetic waves, "they cannot penetrate the strong fields in the atoms." If atoms are essentially void systems, how is it that all bodies are not transparent? He elaborates:

> SIMPLICIO: How could poor me bar the light with my skeleton body, which, so I understand, is so riddled with holes, that a bare thousand-billionth part of it is Something (sic), while the space in between is occupied by fields completely permeable to the blessed rays of light—so to speak non-existent? If that be so, I fail to understand why my standing in front of the window should prevent Sr Salviati from reading.

This is how Schrödinger's Galilean fragment ends, rather abruptly.

It is clear from the manuscript that Schrödinger wanted it to convey as closely as possible the atmosphere of Galileo's original dialogue. Although he implicitly refers to Lorentz, Millikan and Rutherford in his fragment, it would break the historical illusion to mention the three physicists by name. The same holds for the technical terms in the *Fragment*, but only to some extent. At the time of Galileo, the electron was of course unknown and so was the alpha particle, none of which terms appear in the manuscript. Instead of referring to the scattering experiments with alpha particles, Schrödinger chose the old term 'corpuscle' which just means a minute particle of matter. (J.J. Thomson used the term for electrons, but this is evidently not what Schrödinger had in mind.) On the other hand, Schrödinger cannot do without employing terms which would be anachronistic in a seventeenth-century context. He speaks of "volts", "electromagnetic," "electro-static" and "molecules," none of which terms would make sense to Galileo and his contemporaries (for the various terms and their history, see [Kragh 2024]). Another anachronism appears when Salviati says about a sphere of brass that it has "the size of a football."

## Conclusion

Although written by one of the giants of twenty-century physics, Schrödinger's Galilean fragment published in 1955 and republished twenty years later has received almost no attention by physicists and historians of physics. The *Fragment* is not an important work, and it is clearly unimportant from a scientific point of view. Nonetheless, it is interesting from both a biographical and historical perspective. I cautiously suggest, or speculate, that Schrödinger may originally have composed a rough version of the fragment in about 1914, the time to which the dialogue refers. On this assumption the content makes much better sense than it does if it is assumed that he wrote it down in 1955, just a few months before he left Dublin for Vienna. Perhaps Schrödinger scholars can come up with some other explanation of why the *Fragment* is limited to pre-quantum physics.

**Acknowledgments**: My thanks to Martin Gronau, Leibniz-Zentrum für Literatur- und Kulturforschung, Berlin, for critical and inspiring comments.